\documentstyle[graphicx,epsfig,12pt]{article}
\title{{\LARGE\bf High energy photon colliders.}\thanks{Invited talk at the
  International Symposium on New Visions in Laser-Beam Interactions,
  October 11-15, 1999, Tokyo, Metropolitan University Tokyo, Japan. To
  be published in Nucl. Instr. and Meth. B.}  }

\author{Valery Telnov, \\
  {\small\it Budker Institute of Nuclear Physics, 630090 Novosibirsk, 
  Russia}\thanks{email:telnov@inp.nsk.su} \\
  {\small\it and DESY, Germany}} 

\date{}
\begin{document}
\newcommand{\EP}{\mbox{e$^+$}}
\newcommand{\EM}{\mbox{e$^-$}}
\newcommand{\EPEM}{\mbox{e$^+$e$^-$}}
\newcommand{\EMEM}{\mbox{e$^-$e$^-$}}
\newcommand{\EE}{\mbox{ee}}
\newcommand{\GG}{\mbox{$\gamma\gamma$}}
\newcommand{\GP}{\mbox{$\gamma$e$^+$}}
\newcommand{\GE}{\mbox{$\gamma$e}}
\newcommand{\LGE}{\mbox{$L_{\GE}$}}
\newcommand{\LGG}{\mbox{$L_{\GG}$}}
\newcommand{\LEE}{\mbox{$L_{\EE}$}}
\newcommand{\TEV}{\mbox{TeV}}
\newcommand{\WGG}{\mbox{$W_{\gamma\gamma}$}}
\newcommand{\GEV}{\mbox{GeV}}
\newcommand{\EV}{\mbox{eV}}
\newcommand{\CM}{\mbox{cm}}
\newcommand{\M}{\mbox{m}}
\newcommand{\MM}{\mbox{mm}}
\newcommand{\NM}{\mbox{nm}}
\newcommand{\MKM}{\mbox{$\mu$m}}
\newcommand{\E}{\mbox{$\epsilon$}}
\newcommand{\EN}{\mbox{$\epsilon_n$}}
\newcommand{\EI}{\mbox{$\epsilon_i$}}
\newcommand{\ENI}{\mbox{$\epsilon_{ni}$}}
\newcommand{\ENX}{\mbox{$\epsilon_{nx}$}}
\newcommand{\ENY}{\mbox{$\epsilon_{ny}$}}
\newcommand{\EX}{\mbox{$\epsilon_x$}}
\newcommand{\EY}{\mbox{$\epsilon_y$}}
\newcommand{\SEC}{\mbox{s}}
\newcommand{\CMS}{\mbox{cm$^{-2}$s$^{-1}$}}
\newcommand{\MRAD}{\mbox{mrad}}
\newcommand{\IND}{\hspace*{\parindent}}
\newcommand{\beq}{\begin{equation}}
\newcommand{\eeq}{\end{equation}}
\newcommand{\beqn}{\begin{eqnarray}}
\newcommand{\eeqn}{\end{eqnarray}}
\newcommand{\dst}{\displaystyle}
\newcommand{\bm}{\boldmath}
\newcommand{\BX}{\mbox{$\beta_x$}}
\newcommand{\BY}{\mbox{$\beta_y$}}
\newcommand{\BI}{\mbox{$\beta_i$}}
\newcommand{\SX}{\mbox{$\sigma_x$}}
\newcommand{\SY}{\mbox{$\sigma_y$}}
\newcommand{\SZ}{\mbox{$\sigma_z$}}
\newcommand{\SI}{\mbox{$\sigma_i$}}
\newcommand{\SIP}{\mbox{$\sigma_i^{\prime}$}}
\newcommand{\n}{\mbox{$n_f$}}
\maketitle

\begin{abstract}
  
  Using the laser backscattering method at future linear colliders one
  can obtain \GG\ and \GE\ colliding beams (photon colliders) with
  energy and luminosity comparable to that in \EPEM\ collisions.  This
  option has been included in the pre-conceptual designs of linear
  colliders and in work on a Technical Design Report which is in
  progress.  The physics motivation for photon colliders is quite
  clear. The proof of its technical feasibility and the search for the
  best solutions is of first priority now.  A key element of a photon
  collider is a laser with high peak power and repetition rate.  One
  very promising way to overcome this problem is the optical cavity
  approach which is discussed in this paper.  A very high
  $\gamma\gamma$ luminosity could be achieved by further decreasing
  the beam emittances. This will be very challenging. One possible way
  is laser cooling of electron beams. This method is discussed in my
  second talk at this symposium. The solution to the first problem is
  vital for photon colliders and provides an interesting physics
  program. Solution of the second problem makes photon colliders a
  very powerful instrument for study of matter, the best for study of
  many phenomena. How to achieve these goals is the subject of this
  talk.
\end{abstract}

\section{Introduction.}

Fantastic progress in  laser technique makes it possible now to
consider seriously many different applications of lasers in particle
beam physics. I hope that our Symposium (perhaps the first of a series) on New
Visions in Laser-Beam Interactions will be very useful
for progress in this new branch of science.

In this talk I will report on developments in \GG,\GE\ colliders
(shortly Photon Colliders) with energies of about $10^{12}$ eV. The key
element of the Photon Collider is a powerful laser which is used for
production of high energy photons using backward Compton
scattering. Such colliders provide a new unique way for the study of 
matter, similar to \EPEM\ or pp colliders but even better for the study of
some phenomena.

The history of the \GG\ physics and photon colliders is closely
connected with the history of the \EPEM\ colliders. Since 1970
two-photon physics has been actively studied at \EPEM\ storage rings
in collisions of virtual photons.  The spectrum of these photons is
$dn \sim 0.035 d\omega/\omega$, so that \GG\ luminosity was much lower
than that in \EPEM\ collisions.  Nevertheless, these experiments have
provided a lot of new information on the nature of elementary particles.

The maximum energy of \EPEM\ storage rings is limited by  severe
synchrotron radiation. To explore the energy region beyond LEP-II,
linear \EPEM\ colliders (LC) in the range from a few hundred GeV to
about 1.5 TeV are under intense study around the world. Three specific
projects NLC (North American)~\cite{NLC}, TESLA (European)~\cite{TESLA}
and JLC (Asian)~\cite{JLC} have published their pre-conceptual design
reports and  intend to submit full conceptual
design reports in 2001-2002. One team at CERN is working on the
concept of a multi-TeV linear collider (CLIC)~\cite{CLIC} which will be able
extend  the energy range of LC in future.

\begin{figure}[!hbt]
\centering
\vspace*{0.7cm}
\epsfig{file=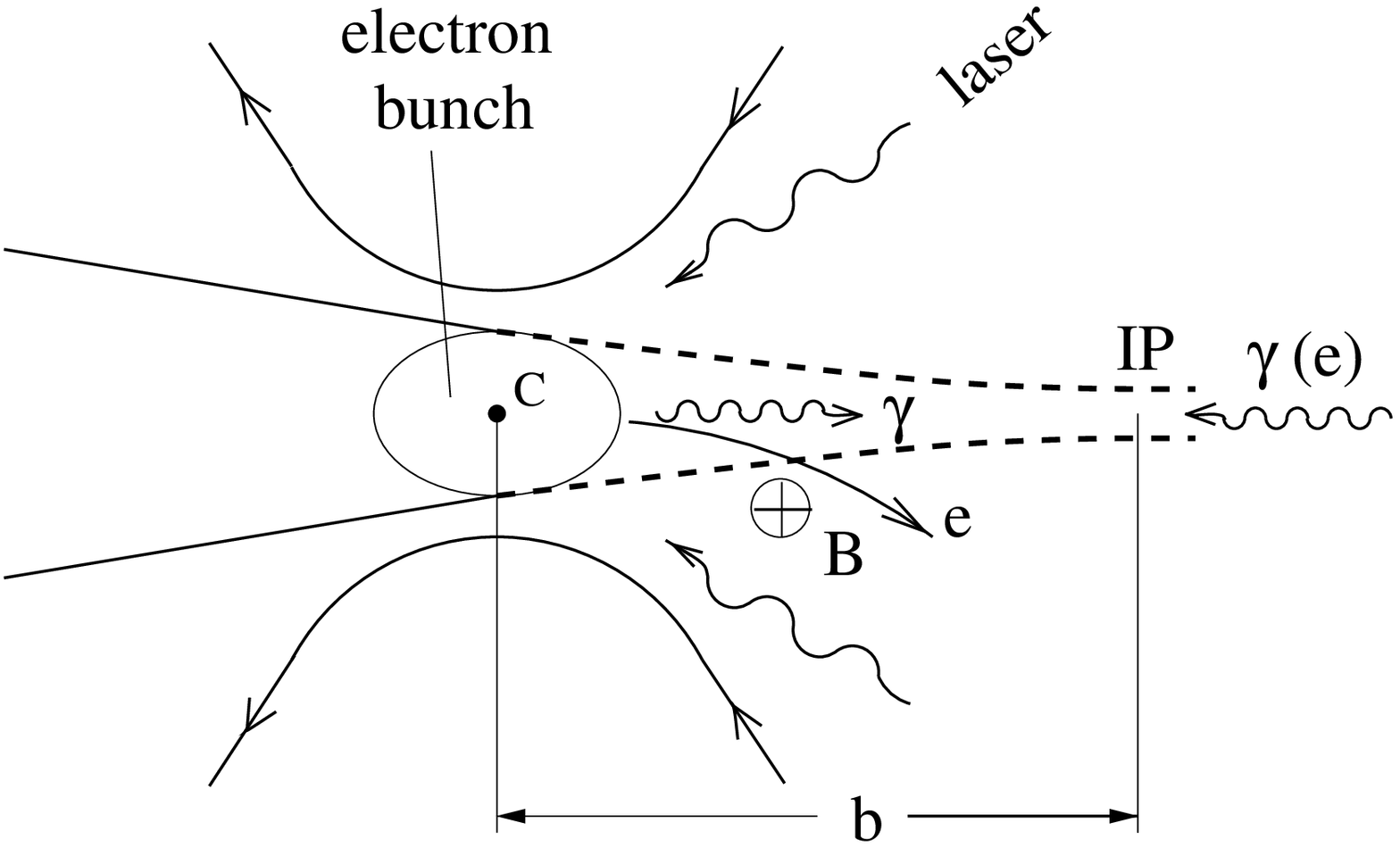,height=9cm,angle=-90}
\vspace*{-0.cm}
\caption{Scheme of  \GG, \GE\ collider.}
\label{ris1}
\end{figure}

Unlike the situation in storage rings, in linear colliders each beam
is used only once. This make it possible to "convert" electrons to
high energy photons to obtain colliding \GG, \GE\ beams. Among various
methods of $e \to \gamma$ conversion the best one is Compton
scattering of the laser light on the high energy electrons.  The basic
scheme of a photon collider is shown in Fig.~\ref{ris1}. Two electron
beams after the final focus system are traveling toward the
interaction point (IP) and at a distance of about 0.1-1 cm from the IP
collide with the focused laser beams.  After scattering, the photons
have an energy close to that of the initial electrons and follow their
direction to the interaction point (IP) (with some small additional
angular spread of the order of $1/\gamma$), where they collide with a
similar counter moving high energy beam or with an electron beam.
With reasonable laser parameters (several Joules flash energy) one can
``convert'' most of the electrons into high energy photons.  The
luminosity of \GG, \GE\ collisions will be of the same order of
magnitude as the ``geometric'' luminosity of the basic $ee$ beams.
Luminosity distributions in \GG\ collisions have the characteristic
peaks near the maximum invariant masses with a typical width about 10
\% (and a few times smaller in \GE\ collisions).  High energy photons
can have various polarizations, which is very advantageous for
experiments.  This idea was proposed by the author and colleagues many
years ago~\cite{GKST81} and has been further developed and discussed
in Refs~\cite{GKST83}-\cite{Tsit1} and many others papers.

The physics at high energy \GG,\GE\ colliders is very rich and no less
interesting than with pp or \EPEM\ collisions (some examples will be
given below). This option has been included in the pre-conceptual
design reports of all LC projects~\cite{NLC}-\cite{JLC}, and work
on the full conceptual design is under way.  

In the present climate of tight HEP budgets we should give very clear
answers to the following questions:

a) can \GG,\GE\ collisions give new physics information in addition to
\EPEM\ collisions that could justify an additional collider cost
($\sim$15\%, second interaction region, including detector)?

b) is it technically feasible?

c) are there enough people for the design and construction of a photon
collider and then exploiting its unique science?
 
Items a) and b) are discussed in the main part of this paper. As for
the last question, the situation is the following.  In the last two
decades, the conception of photon colliders has been developed and
discussed at many workshops. The bibliography on \GG, \GE\ physics now
numbers over 1000 papers, mostly theoretical.  The next phase will
require much wider participation of the experimental community. Now
the work on photon colliders is being continued within the framework of
the Worldwide Study on Physics and Detectors at LC, also  the
International Collaboration on Photon Colliders has recently been initiated.

\section{Physics}

In general, the physics at \EPEM\ and \GG,\GE\ colliders is quite
similar because  the same new particles can be produced. However, the events
are  complimentary, because the cross sections depend
differently on the parameters of the theories. 

If something new is discovered (Higgs, supersymmetry or ...  quantum
gravity with extra dimensions), the nature of these new phenomena will
be better undersood if they are be studied in different reactions.
Some phenomena can best be studied  at photon colliders.
Below I will give several examples.

The second aspect, important for physics study, is the 
luminosity attained by the collider. In the next section it will be shown that
in the current LC designs the \GG\ luminosity in the high energy peak
of the luminosity spectrum is about 20 \% of the \EPEM\ luminosity.
However, if  beams with smaller emittances are used, the \GG\ 
luminosity can be higher than that of \EPEM\ collisions. That is
because in \EPEM\ collisions the luminosity is restricted by the
collision effects (beamstrahlung, instabilities) which are absent in
\GG\ collisions.

\vspace{2mm}
{\it Higgs boson}
\vspace{2mm}

The present "Standard" model, which describes precisely almost
everything at present energies, assumes existence of a very unique
particle, the Higgs boson, which is thought to be responsible for the
origin of particle masses. It is not found yet, but from existing
experimental information it follows that, if it exists, its mass is
about 100--200 GeV, i.e. lays in the region of the next linear colliders.

\begin{figure}[!htb]
\centering
\vspace*{-1.0cm} 
\hspace*{0cm} \epsfig{file=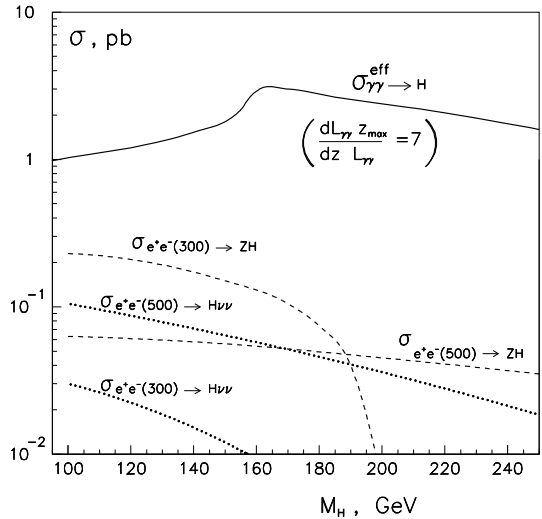,width=10cm,angle=0} 
\vspace*{-1.0cm} 
\caption{Cross sections for the Standard model Higgs in \GG\ and
 \EPEM\ collisions.}
\vspace{0mm}
\label{cross}
\end{figure}

In \GG\ collisions the Higgs boson will be produced as a single
resonance. This process goes via the loop and its cross section is
very sensitive to all heavy (even super-heavy) charged particles.
  The effective cross section is presented in Fig.~\ref{cross}~\cite{ee97}.
Note that here \LGG\ is defined as the \GG\ luminosity at the high
energy luminosity peak ($z=\WGG/2E_e>0.65$ for $x=4.8$) with FWHM
about 15\%. For comparison, the cross sections of the Higgs production
in \EPEM\ collisions are shown.
We see that for $M_H=$ 120--250 GeV the effective cross section in
\GG\ collisions is larger than that in \EPEM\ collisions by a factor
of about 6--30.  If the Higgs is light enough, its width is much less
than the energy spread in \GG\ collisions. It can be detected as a
peak in the invariant mass distribution or can be searched for by energy
scanning using the very sharp ($\sim 1\%$) high energy edge of luminosity
distribution~\cite{ee97}. 
The total number of events in the main decay channels $H \to b\bar b,
WW(W^*), ZZ(Z^*)$ will be several thousands for a typical integrated
luminosity of 10 fb$^{-1}$. The scanning method also enables
the measurement of the Higgs mass with a high precision.

What is most remarkable in this process? The cross section of the
process $\gamma\gamma \to H \to b\bar{b}$ is proportional to
$\Gamma_{\GG}(H)\times Br(H\to b\bar{b}$). The branching ratio
$Br(H\to b\bar{b})$ can be measured with  high precision in \EPEM\ 
collisions in the process with the "tagged" Higgs production: $\EPEM
\to ZH$~\cite{Bataglia}. As a result, one can measure the
$\Gamma_{\GG}(H)$ width at photon colliders with an accuracy better
than 2-3\%~\cite{Jikia},\cite{Melles}. On the other hand, the value of
this two-photon decay width is determined by the sum of contributions
to the loop of all heavy charge particles with masses up to 
infinity.  So, it is a unique way to "see" particles which cannot
be produced at the accelerators directly (maybe never).

The measurement of the Higgs two-photon width reminds me of the
experiment on the measurement of the number of neutrino generations at
LEP. This experiment showned that there are only three light
neutrinos, all of them were known already.  But there could be more.
That would be a great discovery!  Measurement of the Higgs two-photon
width is also some kind of counting of unknown particles.  The Higgs
two-gluon decay width is also sensitive to heavy particles in the
loop, but only to those which have strong interactions (like quarks).
These two measurements together with the $\Gamma_{Z\gamma}(H)$ width,
which could be measured in \GE\ collisions, will allow us to "observe"
and perhaps understand  the nature of invisible heavy charged
particles. This would be a great step forward.

\vspace{2mm}
{\it Charge pair production}
\vspace{2mm}

   The second example is the charged pair production. It could be
   $W^+W^-$ or $t\bar{t}$ pairs or some new, for instance,
   supersymmetric particles.  Cross sections for the production of
   charged scalar, lepton, and top pairs in \GG\ collisions are larger
   than those in \EPEM\ collisions by a factor of approximately 5--10;
   for WW production this factor is even larger, about 10--20. The
   corresponding graphs can be found  elsewhere~\cite{TEL95},\cite{TESLA}.
   
   The cross section of the scalar pair production, predicted in some
   theories, in collision of polarized photons near the threshould, is
   higher than that in \EPEM\ collisions by a factor of 10--20(see
   figs in Refs~\cite{TKEK},\cite{Tfrei}). The cross section in the
   \GG\ collisions near the threshold is very sharp (while in \EPEM\ 
   it contains a factor $\beta^3$) and can be used for measurement of
   particle masses.

Note, that in \EPEM\ collision two charged pairs are produced both via
annihilation diagram with virtual $\gamma$ and $Z$ and also via exchange
diagrams where new particles can  contribute, while in \GG\ 
collisions it is pure QED process which allows the charge of produced particles
to be measured unambiguously.  This is a good example of
complementarity in the study of the same particles in different types of
collisions.

\vspace{2mm}
{\it Accessible masses}
\vspace{2mm}

In \GE\ collisions, charged particle with a mass higher than that in
\EPEM\ collisions can be produced (a heavy charged particle plus a
light neutral), for example, supersymmetric charged particle plus
neutralino or new W boson and neutrino.  \GG\ collisions also provide
higher accessible masses for particles which are produced as a single
resonance in \GG\ collisions (such as the Higgs boson).

\vspace{2mm}
{\it Quantum gravity effects in Extra Dimensions}
\vspace{2mm}

This new theory~\cite{Arkani} suggests a possible explanation of why
gravitation forces are so weak in comparison with electroweak forces.
According to this theory the gravitational forces are as strong as
electroweak forces at small distances in space with extra dimensions
and became weak at large distances due to ``compactification'' of
these extra dimensions.  It turns out that this extravagant theory can
be tested at linear colliders and according to T.Rizzo~\cite{RIZZO}
($\GG\ \to WW$) and K.Cheung~\cite{CHEUNG} ($\GG \to \GG$) photon
colliders are sensitive up to a factor of 2 higher quantum gravity
mass scale than \EPEM\ collisions.

Concluding remark. We have seen that the Higgs and charged pair cross
sections in \GG\ collisions are higher that those in \EPEM collisions
at least by a factor of 5, so, even with 5 times lower \GG\ luminosity
(as it is approximately in current designs) the number of events in
\EPEM\ and \GG\ collisions will be comparable (but physics
complementary). However, the possibility of much larger \GG\ 
luminosity is not excluded, see below.   

\section{Lasers, optics}

The new key element at photon colliders is a powerful laser system
which is used for e$\to \gamma$ conversion.  Lasers with the required
flash energies (several Joules) and pulse duration $\sim$ 1 ps already
exist and are used at several laboratories, the main problem here is
the high repetition rate, about 10--15 kHz.  One very promising way to
overcome this problem is discussed in this paper.  It is an optical
cavity approach, which allows a considerable reduction of the required
peak and average laser power.

\subsection{Requirements for the laser, wave length, flash energy}   
   
The processes in the conversions region: Compton scattering and
several other important phenomena have been considered in detail in
papers~\cite{GKST83},\cite{GKST84},\cite{TEL90},\cite{TEL95},\cite{Monter}
and references therein. There you can find formulae, figures and
explanation of various phenomena in the conversion region as well as
requirements for lasers for photon colliders.  

Laser parameters important for this task are: laser flash energy,
duration of laser pulse, wave length and repetition rate.  The
required wave length follows from the kinematics of Compton
scattering~\cite{GKST83}. In the conversion region a laser photon with
the energy $\omega_0$ scatters at a small collision angle $\alpha_0$
on a high energy electron with the energy $ E_0$.  The maximum energy
of scattered photons (in direction of electrons)
\begin{equation}
\omega_m=\frac{x}{x+1}E_0; \;\;\;\;
x=\frac{4E_0\omega_0\cos^2{\alpha/2}}{m^2c^4}
 \simeq 15.3\left[\frac{E_0}{\TEV}\right]
\left[\frac{\omega_0}{eV}\right].
\end{equation}
For example: $E_0$ =250\,\, GeV, $\omega_0 =1.17$ eV
($\lambda=1.06$ \MKM) (Nd:Glass laser) $\Rightarrow$ $x=4.5$ and
$\omega/E_0 = 0.82$.  The energy of the backscattered photons grows
with increasing $x$.  However, at $x > 4.8$ the high energy photons
are lost due to \EPEM\ creation in the collisions with laser
photons~\cite{TEL90},\cite{TEL95}. The maximum conversion coefficient
(effective) at $x\sim 10$ is about 0.33 while at $x < 4.8$ it is about
0.65 (one conversion length). The luminosity in the first case will be
smaller by a factor of 4.  Detailed study of dependence of the maximum
\GG\ luminosity and monochromaticity on $x$ can be found
elsewhere~\cite{TEL90}.

In the laser focus at photon colliders the field is so strong that
multi-photon processes can take place, for example, the electron can
scatter simultaneously on several laser photons. It is preferable to
work in a regime where these effects are small enough, because the
shape of the photon spectrum in this case is sharper.  Sometimes
strong fields can be useful. Due to transverse motion of electrons in
the laser wave the effective electron mass is increased and the
threshold of \EPEM\ production is shifted to the higher beam energies,
a factor of 1.5--2 is possible without special problems ``simply'' by
adding a laser power.  For some tasks, such as the energy scanning of
the low mass Higgs, the luminosity spectrum should be very sharp, that
is only possible when multi-photon effects are small.

From all this it follows that an existing powerful Terawatt solid state
laser with the wave length about 1 \MKM\ can be used for photon colliders up
to c.m.s. energies about 1 TeV. For low energy colliders (for study of
the low mass Higgs, for instance), the doubling  of the
laser frequency may be useful, this can be done with high efficiency.

In the calculation of the required flash energy one has to take into
account the natural ``diffraction'' emittance of the laser
beam~\cite{GKST83}, the maximum allowed value of the field strength
(characterized by the parameter $\xi^2 =
e^2\bar{B^2}\lambda^2/m^2c^4$) \cite{TEL90}, \cite{TEL95} and the laser
spot size at the conversion point which should be larger than that of
the electron beam. In the collision scheme with the
"crab-crossing"~\footnote{The crab crossing scheme for beam
  collisions~\cite{PALMER} is obligatory in photon colliders for the
  removal of disrupted beams~\cite{TEL90}. In this scheme the electron
  bunches are collided with crossing angle $\alpha_c$.  To preserve the
  luminosity the electron bunches are tilted (using an RF cavity) with
  respect to the direction of the beam motion on the angle
  $\alpha_c/2$. The required $\alpha_c$ for the  projects considered is
  about 30 mrad~\cite{TESLA}.}  the electron beam is tilted in respect
to the direction of motion that creates an additional effective
transverse beam size $\sigma_x = \sigma_z\alpha_c/2$.  The result of
MC simulation of $k^2$ ($k$ is the conversion coefficient, $k^2$ is
proportional to the \GG\ luminosity) for the electron bunch length
$\sigma_z= 0.3$ mm (TESLA project) as a function of the flash energy
and parameter $\xi^2$ (in the center of the laser bunch) are shown in
figs. 3 and 4.
\begin{figure}[!htb]
 
\vspace{-0.9cm}
 
\hspace*{0cm}\begin{minipage}[b]{0.45\linewidth}
\centering
\vspace*{-0.cm}
\hspace*{-0.5cm} \epsfig{file=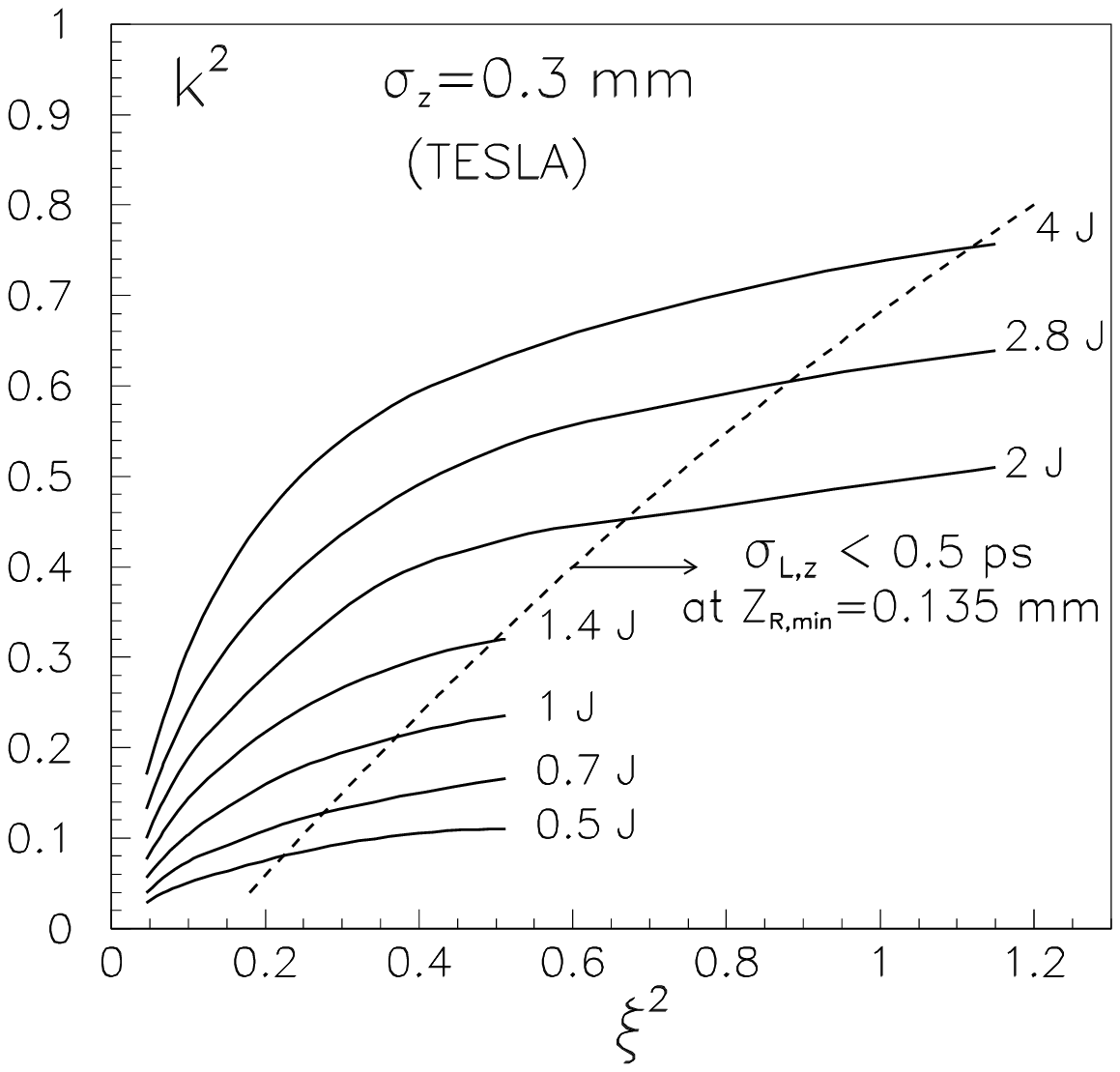,width=8.2cm,angle=0}
 
\vspace{-0.8cm}
\label{tes} 
\caption{Square of the conversion probability  luminosity as a function 
  of the laser flash energies for various the values of the parameter
  $\xi^2$. Electron beams pass through the holes in the mirrors. See
  comments in the text.}
\end{minipage}%
\hspace*{1.5cm} \begin{minipage}[b]{0.45\linewidth} \centering
 
\vspace*{-2cm}

\hspace*{-1.cm} \epsfig{file=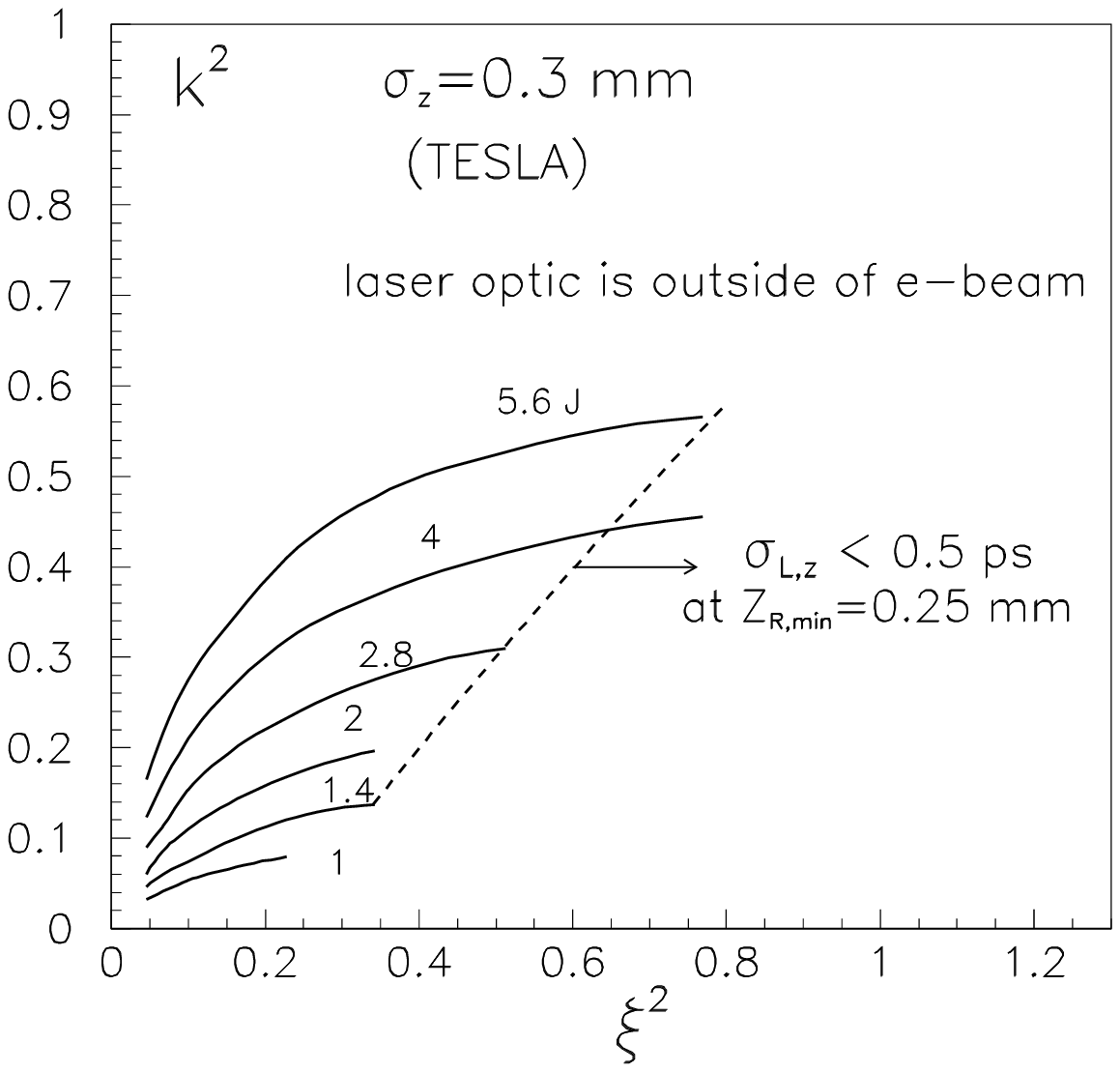,width=8.2cm,angle=0}
 
\vspace{-.7cm}
\label{tes2r}
\caption{Same as on the previous figure, but the mirror system is 
situated outside  the electron beam trajectories.  
\vspace{1.3cm}}
\end{minipage}
\vspace{-0.3cm}
\end{figure}

In summary: the required laser flash energy is about 3--5 Joules, which is
quite reasonable. However, the LC have a repetition rate of about
10--15 kHz, so the average power of the laser system should be 
about 50 kW. One  possible solution is the multi-laser system which
combines pulses into one train using Pockels cells~\cite{NLC}.
However, such a system will be  very expensive~\cite{Perry}.

\subsection{Multi-pass laser systems}

To overcome the ``repetition rate'' problem it is quite natural to
consider a laser system where one laser bunch is used for e$\to
\gamma$ conversion many times. Indeed, one Joule laser flash contains about
$10^{19}$ laser photons and only $10^{10}-10^{11}$ photons are knocked out in
the collision with one electron bunch. 

The simplest solution is to trap the laser pulse to some optical loop and
use it many times.~\cite{NLC} In such a system the laser pulse enters
via the film polarizer and then is trapped using Pockels cells and
polarization rotating plates.  Unfortunately, such a system will not
work with Terawatt laser pulses due to a self-focusing effect.

Fortunately, there is one way to ``create'' a powerful laser pulse in
the optical ``trap'' without any material inside. This very promising
technique is discussed below.

\subsection{Laser pulse stacking in an ``external'' optical cavity.} 

Shortly, the method is the following. Using the train of low energy
laser pulses one can create in the external passive cavity
(with one mirror having some small transparency) an optical pulse of
the same duration but with much higher energy (pulse stacking). This
pulse circulates many times in the cavity each time colliding with
electron bunches passing the center of the cavity.

The idea of pulse stacking is simple but not trivial and not well
known in the HEP community (and even to laser experts, though it is as old
as the Fabry-Perot interferometer). This method is used now in several
experiments on detection of gravitation waves. It was mentioned also
in NLC ZDR~\cite{NLC} though without analysis and further development.
   In my opinion, pulse stacking is very natural for photon colliders
and allows not only to build a relatively cheap laser system for
$e\to\gamma$ conversion but gives us the practical way for realization of the
laser cooling, i.e. opens up the way to ultimate luminosities of photon
colliders. 

\begin{figure}[!htb]
\centering
\vspace*{0.2cm} 
\hspace*{-0.2cm} \epsfig{file=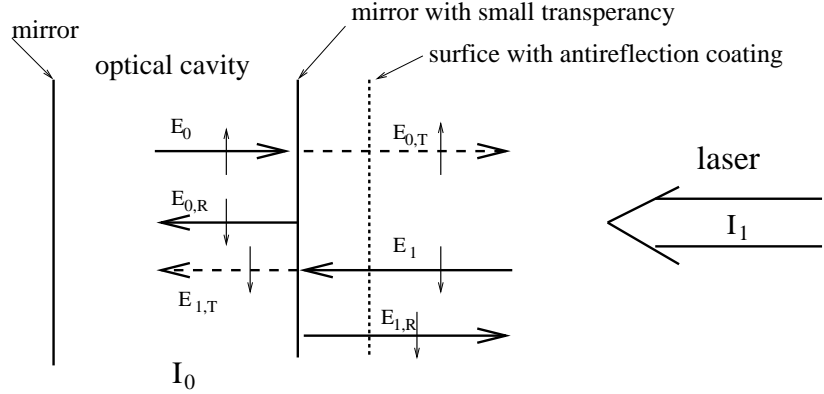,width=11cm,angle=0} 
\vspace*{-0.cm} 
\caption{Principle of pulse stacking in an external optical cavity.}
\vspace{2mm}
\label{cavity}
\vspace{-2mm}
\end{figure}

As this is very important for photon colliders, let me consider this
method in more detail~\cite{Tfrei}. The principle of pulse stacking is
shown in Fig.\ref{cavity}.
The secret consists in the following. There is a well known optical
theorem: at any surface, the reflection coefficients for light coming
from one and the other sides have opposite signs. In our case, this means
that light from the laser entering through semi-transparent mirror into
the cavity interferes with reflected light inside the cavity {\bf
constructively}, while the light leaking from the cavity interferes
with the reflected laser light {\bf destructively}. Namely, this fact produces
asymmetry between cavity and space outside the cavity!  

Let R be the reflection coefficient, T the transparency coefficient
and $\delta$ the passive losses in the right mirror. From the energy
conservation $R+T+\delta =1$. Let $E_1$ and $E_0$ be the amplitudes
of the laser field and the field inside the cavity. In equilibrium,
$E_0= E_{0,R} + E_{1,T}$. 
Taking into account that $E_{0,R}=E_0\sqrt{R}$, $E_{1,T}=E_1\sqrt{T}$ and 
$\sqrt{R}\sim 1-T/2-\delta/2$ for $R\approx 1$ we obtain
$E_0^2/E_1^2=4T/(T+\delta)^2.$ 
The maximum ratio of intensities is obtained at $T=\delta$, then 
$I_0/I_1=1/\delta \approx Q$,
where $Q$ is the quality factor of the optical cavity.  Even with two
metal mirrors inside the cavity, one can hope to get a gain factor of about
50--100; with multi-layer mirrors it can reach $10^5$. ILC(TESLA)
colliders have 120(2800) electron bunches in the train, so the factor
100(1000) would be perfect for our goal, but even the factor of ten
means a drastic reduction of the cost.

   Obtaining of high gains requires a very good stabilization of cavity
size: $\delta L \sim \lambda/4\pi Q$, laser wave length: $\delta
\lambda/\lambda \sim \lambda/4\pi QL$ and distance between the laser
and the cavity: $\delta s \sim\lambda/4\pi$. Otherwise, the  condition of
constructive interference will  not be fulfilled. Besides, the
frequency spectrum of the laser should coincide with the cavity modes,
that is automatically fulfilled when the ratio of the cavity length and
that of the laser oscillator is equal to an integer number 1, 2, 3... . 

For $\lambda = 1\;\mu m$ and $Q=100$, the stability of the cavity
length should be about $10^{-7}$ cm. In the LIGO experiment
 on detection of gravitational waves which uses 
similar techniques with $L\sim 4$ km and $Q\sim 10^5$ the expected
sensitivity is about $10^{-16}$ cm.  In comparison with this project
our goal seems to be very realistic.

      In HEP literature I have found only one reference on pulse
stacking of short pulses ($\sim 1$ ps) generated by FEL~\cite{HAAR}
with the wave length of 5 $\mu$m. They observed pulses in the cavity
with 70 times the energy of the incident FEL pulses, though no long
term stabilization was done.
   
\begin{figure}[!htb]
\centering
\vspace*{-0.3cm} 
\hspace*{-0.4cm} \epsfig{file=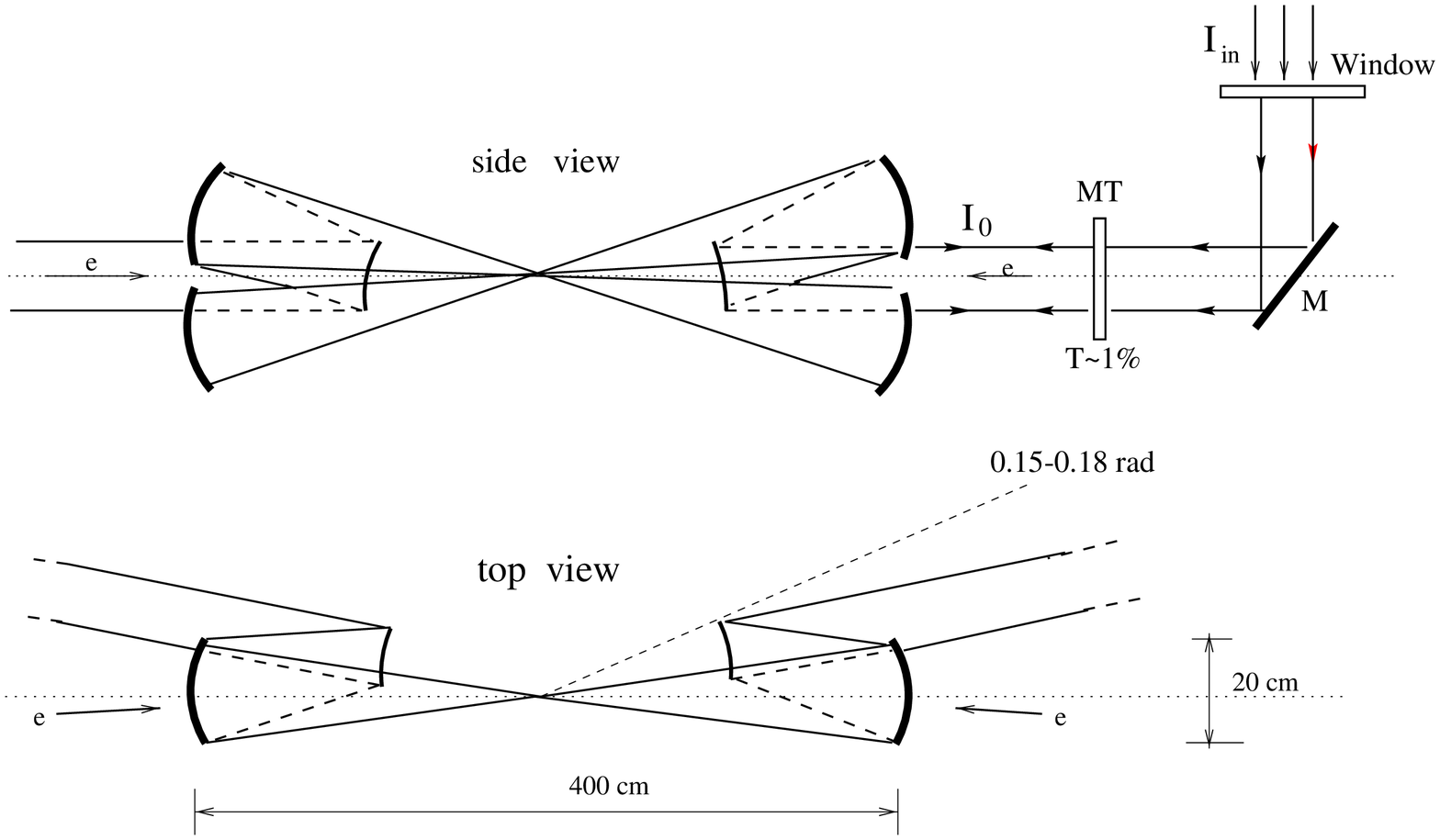,width=12cm,angle=0} 
\vspace*{-0.0cm} 
\caption{Possible scheme of optics at the IR.}
\vspace{0mm}
\label{optics}
\vspace{-6mm}
\end{figure}

Possible layout of the optics at the interaction region scheme is
shown in Fig.\ref{optics}. In this variant, there are two optical
cavities (one for each colliding electron beam) placed outside the
electron beams.  Another possible variant has only one cavity common
for both electron beams. In this case, it is also possible to arrange
two conversion points separated by the distance of several millimeters
(as it is required for photon colliders), though the distribution of
the field in the cavity is not completely stable in this case (though
it may be sufficient for not too large a Q and , it can be made stable
in more complicated optical system). Also, mirrors should have holes
for electron beams (which does not change the Q factor of the cavity
too much). The variant presented in fig.\ref{optics} is simpler though
it requires a factor of 2 higher flash energy.

\section{Luminosity of photon colliders in current designs.}
Some results of simulation of \GG\ collisions at TESLA, ILC (converged
NLC and JLC) and CLIC are presented below in Table~\ref{table1}. Beam
parameters were taken the same as those in \EPEM\ collisions with the
exception of the horizontal beta function at the IP which is taken (quite
conservatively) equal to 2 mm for all cases, that is several times
smaller than that in \EPEM\ collisions due to the absence of
beamstrahlung.  The conversion point(CP) is situated at distance
$b=\gamma\sigma_y$. It is assumed that electron beams have 85\%
longitudinal polarization and laser photons have 100\% circular
polarization.

%\vspace{-0.3cm}
%{\setlength{\tabcolsep}{2.mm}
%{\footnotesize
%{
\begin{table}[hbt]
\caption{Parameters of  \GG\ colliders based on TESLA, ILC (NLC/JLC)}
\vspace{.2cm}
\renewcommand{\arraystretch}{1}
\begin{center}
\hspace*{-2.3mm}\begin{tabular}{l c c c c } \hline
& T(500) & I(500)   &  T(800) & I(1000)  \\ \hline \hline 
\multicolumn{5}{c}{ no deflection, $b=\gamma \sigma_y$, $x=4.6$} \\ \hline
$N/10^{10}$& 2. & 0.95 & 1.4 & 0.95 \\  
$\sigma_{z}$, mm & 0.4 & 0.12 & 0.3 & 0.12 \\  
$f_{rep}$, Hz& 5 & 120 & 3& 120 \\
$n_b/train$ & 2820 & 95 &4500 & 95 \\
$f_{rep}\times n_b$, kHz& 14.1 & 11.4& 13.5 & 11.4 \\
$\Delta t_b$, ns  & 337 & 2.8 & 189 & 2.8  \\ 
$\gamma \epsilon_{x,y}/10^{-6}$,m$\cdot$rad & $10/0.03$ & $5/0.1$ 
&  $8/0.01$ & $5/0.1$ \\
$\beta_{x,y}$,mm at IP& $2/0.4$ & $2/0.12$  &
$2/0.3$& $2/0.16$  \\
$\sigma_{x,y}$,nm& $200/5$ & $140/5$ & 
$140/2$ & $100/4$ \\  
b, mm & 2.4 & 2.4  & 1.5 & 4  \\
$L(geom),\,\,\,  10^{33}$& 48 & 12 & 75 & 20 \\  
$\LGG (z>0.65), 10^{33} $ & 4.5 & 1.1  & 7.2 & 1.75  \\
$\LGE (z>0.65), 10^{33}$ & 6.6 & 2.6 & 8  & 4.2  \\
$\LEE, 10^{33}$ & 1.2 & 1.2  & 1.1 & 1.8 \\
$\theta_x/\theta{_y},_{max}$, mrad ~ & 5.8/6.5 & 6.5/6.9 & 
 4.6/5 & 4.6/5.3  \\ \hline
\vspace{-5.mm}
\end{tabular}
\end{center}
\label{table1}
\end{table}
%}}

We see that the \GG\ luminosity in the hard part of the spectrum $\LGG
(z>0.65)\sim 0.1L(geom)$, numerically it is about $(1/6)L_{\EPEM}$.
\footnote{this is because a) $L_{\EPEM} \sim 1.5L_{geom}$, factor 1.5
  (roughly) is due to the pinch effect: b) $L_{geom}$ in the case of
  photon colliders is larger than that in \EPEM\ collisions by a
  factor about 2.5 (in the current projects) due to the smaller
  $\beta$-function} Note, that the coefficient $1/6$ is not a
fundamental constant.  The \GG\ luminosity in these projects is
determined only by ``geometric'' ee-luminosity. With some new low
emittance electron sources or with laser cooling of electron beams
after the damping ring (or photo-guns) one can get, in principle,
$\LGG (z>0.65) > L_{\EPEM}$.  The limitations and technical
feasibility are discussed in the next section.  In addition to the
\GG\ collisions, there is considerable \GE\ luminosity (see table) and
it is possible to study \GE\ interactions simultaneously with \GG\ 
collisions.
  
\begin{figure}[!htb]
\centering
\vspace*{-1.4cm} 
\hspace*{-0.6cm} \epsfig{file=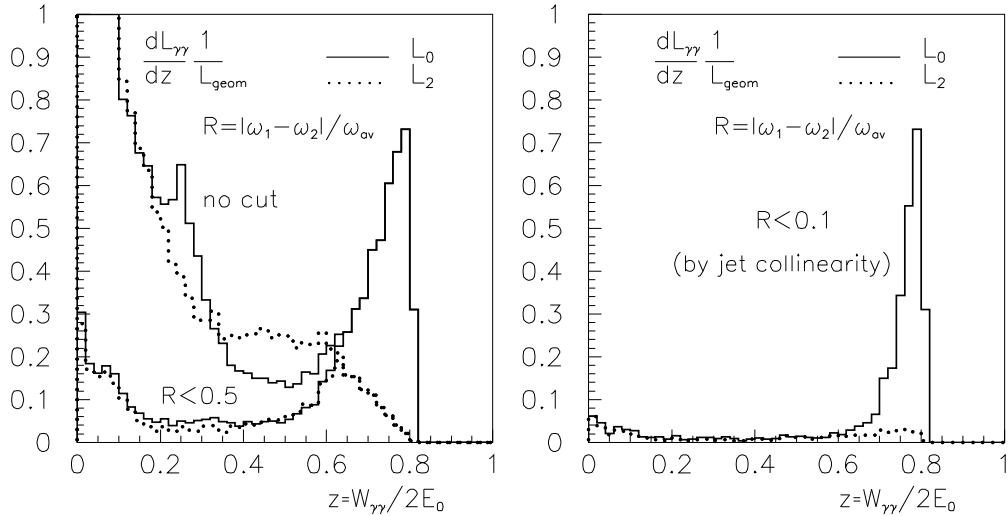,width=17cm,angle=0} 
\vspace{-2.cm} 
\caption{\GG\ luminosity spectra at TESLA(500) for parameters
presented in Table 1. Solid line for total helicity of two photons 0
and dotted line for total helicity 2. Upper curves without cuts, two
lower pairs of curves  have cut on the relative difference
of the photon energy. See comments in the text.}
\label{TeslaR}
\end{figure}

   The normalized \GG\ luminosity spectra for a 0.5 TeV TESLA are
   shown in Fig.\ref{TeslaR}(left).
The luminosity spectrum is decomposed into two parts, with the total
helicity of two photons 0 and 2. We see that in the high energy part
of the luminosity spectra photons have a high degree of polarization,
which is very important for many experiments.  In addition to the high
energy peak, there is a factor 5--8 larger low energy luminosity. It
is produced by photons after multiple Compton scattering and
beamstrahlung photons. Fortunately, these events have a large boost and
can be easily distinguished from the central high energy events.  In
the same Fig.\ref{TeslaR}(left) you can see the same spectrum with an
additional ``soft'' cut on the longitudinal momentum of the produced
system which suppresses low energy luminosity to a negligible level.

Fig.\ref{TeslaR} (right) shows the same spectrum with a stronger cut
on the longitudinal momentum. In this case, the spectrum has a nice
peak with FWHM about 7.5\%. On first sight such cut is somewhat
artificial because one can directly select events with high invariant
masses and the minimum width of the invariant mass distribution depends
only on the detector resolution. However, there is a very important
example when one can obtain a ``collider resolution'' somewhat better
than the ``detector resolution''; this is the
case of only two jets in the event when one can restrict the
longitudinal momentum of the produced system using the acollinearity
angle between jets ($H\to b\bar b, \tau\tau$, for example).

 A  similar table and distributions for the photon collider on the
c.m.s. energy 130 GeV (Higgs collider) can be found in
ref.\cite{TKEK}.

\section{Ultimate \GG, \GE\  luminosities }
There is only one collision effect restricting the \GG\ luminosity,
that is a process of  coherent pair creation when the high energy
photon is converted into an \EPEM\ pair in the strong field of the opposing
electron beam~\cite{CHTEL},\cite{TEL90},\cite{TEL95}.  It becomes more
important at larger collider energies or(and) very short bunches. 
Detailed analysis of ultimate luminosities at photon colliders was
done in the ref.~\cite{TSB2}.

\begin{figure}[!htb]
\centering
\vspace*{-0.7cm} 
\hspace*{-0.7cm} \epsfig{file=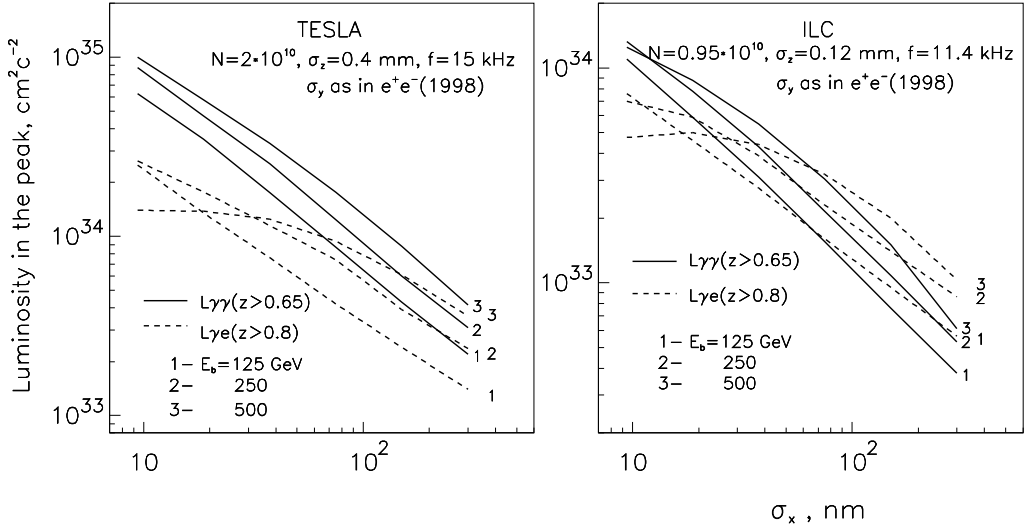,width=15cm,angle=0} 
\vspace*{-1.3cm} 
\caption{Dependence of \GG\ and \GE\ luminosities in the high energy
peak on the horizontal beam size for TESLA and ILC at various
energies. See also comments in the text.}
\vspace{0mm}
\label{sigmax}
\vspace{1mm}
\end{figure}

In the current projects the \GG\ luminosities are determined by the
``geometric'' luminosity of the electron beams. Having electron beams
with smaller emittances one can obtain a much higher \GG\ 
luminosity~\cite{TSB2}. Below are results of the simulation with the
code which takes into account all main processes in beam-beam
interactions~\cite{TEL95}. Fig.\ref{sigmax} shows dependence of the
\GG\ (solid curves) and \GE\ (dashed curves) luminosities on the
horizontal beam size.
 The vertical emittance is taken as in TESLA(500), ILC(500)
projects (see Table \ref{table1}). The horizontal beam size was varied
by change of horizontal beam emittance keeping the horizontal beta
function at the IP constant and equal to 2 mm.

One can see that all curves for \GG\ luminosity follow their natural
behavior: $\L\propto 1/\sigma_x$, with the exception of ILC at
$2E_0=1$ GeV where at small $\sigma_x$ the effect of coherent pair
creation is seen.\footnote{This curve has also some "bend" at large
  $\sigma_x$ that is connected with synchrotron radiation in quads
  (Oide effect) due to a large horizontal emittance. One can avoid
  this effect by taking larger $\beta_x$ and smaller \ENX.}  This
means that at the same collider the \GG\ luminosity can be increased
by decreasing the horizontal beam size (see table 1) at least by one
order ($\sigma_x < 10$ nm is difficult due to some effects connected
with the crab crossing).

Additional increase of \GG\ luminosity by a factor about 3 (TESLA),
7(ILC) can be obtained by a further decrease of the vertical
emittance~\cite{TKEK}. So, using beams with smaller emittances, the
\GG\ luminosity at TESLA, ILC can be increase by almost 2 orders of
magnitude. However, even with one order improvement, the number of
``interesting'' events (the Higgs, charged pairs) at photon colliders
will be larger than that in \EPEM\ collisions by about one order. This
is a nice goal and motivation for photon colliders.

  In \GE\ collision (Fig.\ref{sigmax}, dashed curves), the behavior of
the luminosity on $\sigma_x$ is different due to additional collision
effects: beams repulsion and beamstrahlung. As a result, the
luminosity in the high energy peak is not proportional to the
``geometric''  luminosity.

There are several ways of decreasing the transverse beam emittances
(their product): optimization of storage rings with long wigglers,
development of low-emittance RF (or pulsed photo-guns) with merging many
beams with low charge and emittance.  Here some progress is certainly
possible.  Moreover, there is one method which allows further decrease
of beam cross sections by two orders of magnitude in comparison with current
designs, it is a laser cooling ~\cite{TSB1},\cite{Monter}. This method
is discused in my second talk at this Symposium.

Other important aspects for photon colliders are removal of disrupted
beams and backgrounds. Discussion of these problems can be found
elsewhere~\cite{TEL90}, \cite{TESLA}, \cite{Tsit1}.

\section{Conclusion}

The physics program for photon \GG,\GE\ colliders is very interesting
and the additional cost of the second interaction region is certainly
justified.

There are no show-stoppers. All processes in the conversion and
interaction regions and the limitations of attainable luminosity are
well understood. There are ideas on laser and optical scheme designs.
However, much remains to be done in terms of detailed studies and
experimental tests.

Special effort is required for the development of the laser and optics
which are the key elements of photon colliders. The present laser
technology has, in principle, all elements needed for photon
colliders, the development of a practical scheme is the most pressing
task now.  One of the most promising methods is the optical cavity
approach which allows a considerable reduction of the required peak
and average laser power.  A reduction of one order of magnitude is
already sufficient, but for the TESLA collider with a large number of
bunches in a train and large spacing between the bunches one can think
about 2--3 orders, though this may be difficult due to other reasons.

The \GG\ luminosity at photon colliders with  energy below 
one TeV can be higher than that in \EPEM\ collisions, typical cross
sections are also several times higher, so one could consider an 
X-factory (X = Higgs, W, etc.). The main problem here is the generation of
polarized electron beams with very small emittances (products of
transverse emittances). Optimization of damping rings and development of
low emittance multi-gun RF sources is the first step in this
direction.  The second step requires new technologies. The laser
cooling of electron beams is one  possible way of achieving
ultimate \GG\ luminosity. Realization of this method depends on the
progress of  Laser Technology, especially promising is the method
of the storage (stacking) of laser pulses in an optical cavity.

Dear participants of the Symposium on New Vision in Laser Beam
Interactions, and all laser experts, there is a possibility  to build 
a unique instruments for study of the matter in the next decade: The High
Energy Photon Collider. The development of the required laser systems
is a very challenging task, we need your knowledge, experience and
talent, join us in this exciting undertaking!

\end{document}